\documentstyle[epsf,onecolumn]{mn}
\include{mac}
\begin{document}
\title[PW Vul 84]
{Abundance analysis of the slow nova PW Vulpeculae 1984}
\author[Schwarz et al.]
{Greg J. Schwarz,$^1$
S. Starrfield,$^1$
Steven N. Shore,$^2$
and Peter H. Hauschildt$^3$ \\
$^1$Department of Physics and Astronomy,
Arizona State University, Tempe, AZ 85287-1504 \\
E-Mail: \tt schwarz@hydro.la.asu.edu, 
\tt sumner.starrfield@asu.edu \\
$^2$Department of Physics and Astronomy,
Indiana University South Bend,
1700 Mishawaka Ave, South Bend, IN 46634-7111 \\
E-Mail: \tt sshore@paladin.iusb.edu \\
$^3$Department of Physics and Astronomy,
University of Georgia, Athens, GA 30602-2451 \\
E-Mail: \tt yeti@hal.physast.uga.edu \\}

\maketitle

\begin{abstract}

We determine the elemental abundances for the ejecta of the 
slow nova PW Vul 1984.
Our technique uses a minimization of the emission line fits of a 
photoionization model to available ultraviolet, optical and 
infrared spectra.  We find the following abundances (by number) with
respect to solar: He/H = 1.0 $\pm$ 0.4, C/H = 7.0 $^{+7}_{-4}$,
N/H = 85 $^{+59}_{-41}$ and O/H = 6 $^{+7}_{-2}$.  In addition, there
is weak evidence for solar Ne and Mg and twice solar Fe.
Previous studies (Saizar et al. 1991 and Andre\"{a} et al. 1991, 1994) 
of PW Vul have yielded considerable differences in their 
derived elemental abundances for the ejecta.  Our abundances fall 
in between the previous studies.  To explain the discrepant abundances, we 
analyze in detail the data and methods used to obtain the previous results.  
The abundances of Saizar et al. (1991) are significantly smaller then
our values because of the lower electron temperature used by 
Saizar et al. in deriving elemental abundances from ion abundances.  
Andre\"{a} et al. (1991) used an ionization correction
method to obtain their abundances and verified their results 
with a photoionization model (Andre\"{a} et al. 1994).  Our analysis
of their data shows that the absolute fluxes of the optical emission lines
used by Andre\"{a} are underestimated by 15\% leading to a factor of 2 
increase in their derived abundances.  We also find the photoionization
model used by Andre\"{a} et al. (1994) predicts 2 times more carbon than
the photoionization code we used even when fitting the same data with 
similar model parameters.

\end{abstract}

\begin{keywords}

stars: abundances -- stars: individual -- stars: novae.

\end{keywords}

\section{Introduction}

The slow novae PW Vul 1984  
was discovered on 27.7 July UT by Wakuda (Kosai 1984).  It
reached maximum light of 6.3 mag in the V band on 4.1 August 1984.  
The light curve showed strong oscillations of 1-2 magnitudes 
on the decline (Robb and Scarfe 1995) making a determination 
of t$_{2,3}$ (the time to fall 2 or 3 magnitudes from 
maximum) uncertain.   The light curve of PW Vul was similar to
that of the proto-typical dust forming nova, DQ Her, however, infrared
photometry of PW Vul did not indicate significant dust
formation (Gehrz et al. 1988).

The early lightcurve showed strong oscillations making the 
distance determination from a standard rate-of-decline vs maximum 
magnitude relationship problematic. 
Gehrz et al. (1988) derived a distance of 6.5 kpc from the change 
in the blackbody angular radius of the photosphere 
{\it but assumed no interstellar extinction}.  
Evans et al. (1990) showed that assuming a realistic extinction of
E(B-V)=0.5 increased the angular diameter and reduced Gehrz's 
distance to 3.3 kpc.   Evans et al. (1990) assumed
that PW Vul was emitting at the Eddington luminosity 
of a solar mass star at maximum light and that the
absolute magnitude at maximum was the same as DQ Her
to find 2.3 kpc and 2.6 kpc respectively assuming an E(B-V) = 0.5.  
Ringwald $\&$ Naylor (1996) obtained
1.6 $\pm$ 0.2 kpc from expansion parallax measurements in H$\alpha$ 
taken nine years after outburst.  An average of the last four
distance estimates gives 2.5 $\pm$ 1 kpc as our best estimate
of the distance to PW Vul.

PW Vul has been observed in most wavelength regimes at various times
during the outburst.
Infrared spectra taken early in the outburst showed strong emission 
lines of neutral carbon, nitrogen and oxygen (Evans et al. 1990, 
Williams et al. 1996).  Coronal lines of [S\,{\sc VI}] (1.96 $\mu$m),
[Mg\,{\sc VIII}] (3.02 $\mu$m), [Al \,{\sc VI}] (3.65 $\mu$m) and later
Si \,{\sc VII}] (2.47 $\mu$m) and [Al\,{\sc V}] (2.90 $\mu$m) 
slowly replaced the CNO lines as the 
infrared spectrum evolved (Williams et al. 1996).  Optical spectra
taken during the nebular phase showed strong emission lines of 
[O\,{\sc III}] (5007, 4959 {\AA}), [N\,{\sc II}] (5755 {\AA})
and Balmer hydrogen
(Kenyon $\&$ Wade [1986; hereafter KW], Andre\"{a} et al. [1991 and 1994; 
hereafter A91 and A94], Saizar et al. [1991; hereafter S91]).  
The International Ultraviolet Explorer (IUE) satellite obtained spectra of 
PW Vul from the earliest phases of the outburst to well into the nebular
phase when the ejecta had become optically thin.  The X-ray satellite,
EXOSAT, observed an increase in the X-ray flux over the first year
of the outburst (\"Ogelman et al. 1987).

S91 and A91 determined elemental abundances from nebular spectra.  
S91 and A91 both utilized UV and optical spectra
to determine {\it ionic} abundances of individual ions but differed
in how they computed the {\it elemental} abundances.  As a result,
S91 found a factor of 4.1, 8.5, and 33.3 times
{\it less} nitrogen, oxygen, and carbon, respectively, then A91.
This is an important discrepancy and one that could easily call the
determination of all nova abundances into question.  For this 
reason, in this paper, we re-examine the available 
data and determine new elemental abundances for PW Vul.  We use 
contemporaneous UV, optical, and infrared spectra in our modeling
and use statistically robust minimization techniques to determine the 
photoionization model which predicts the best fit to the data.  
In addition, we critically examine the methods used by A91, A94 and S91  
and suggest an explanation for the differences in their abundances.

\section{Data calibration and reddening}

In order to accurately model the abundances of nova 
ejecta, the spectra must be contemporaneous, flux calibrated and 
dereddened.  We used ultraviolet, optical and infrared spectra 
obtained within a 2 week period during July 1985 because this is the 
largest amount of data available with the smallest time spread.
We are justified in using data over this time span because the
spectral development of PW Vul was slow enough so that even though
the line fluxes changed, the {\it line flux ratios} were 
essentially the same during the months of June and July 1985.
The ultraviolet was represented by the IUE spectrum SWP26342 taken on 
3.09 July 1985.  Unfortunately, there are no LWP observations on the
same day but we used IUE spectrum LWP6264 (obtained on 24.6 June 1985)
and spectrum LWP6408 (obtained on 16.7 July 1985) to 
estimate a Mg\,{\sc ii} (2800 {\AA}) line strength for 3.09 July.
A Gaussian fit was used to determine
the integrated UV emission line fluxes.  The primary source of uncertainty
was in the determination of the continuum level and multiple
Gaussian fits of each UV emission line provided an
estimate of this uncertainty.  An additional 10\% was included in 
the uncertainty to account for the accuracy of the absolute 
flux determination of IUE (Bohlin 1980).

We used A91's spectra obtained during 1-6 July 1985 for the optical range.
To assess the quality of A91's spectra, we compared
them with the optical spectra published by KW from
2 June 1985 (their Table 1) and
S91's 25 September 1985 (their Table 3) spectrum.  The ratios of
the strongest optical lines to H$\beta$ in A91 were consistent
with similar ratios in KW and S91.  A91 places each emission line in an
error class of 10-20\%, 20-50\% or greater than 50\%.  Lacking any more 
information about the error classes of A91, we set the
uncertainty of each optical emission line to the average value of
its assigned error class.  In addition, we included the
infrared Paschen lines Pa$\gamma$ and Pa$\beta$ and He\,{\sc i} 
(1.083 $\mu$m) taken on 21 June 1985 (Williams et al. 1996).  Williams
et al. estimate uncertainties of 10\% for Pa$\beta$ and He\,{\sc i} 1.08
$\mu$m and 20\% for Pa$\gamma$.  Table 1 contains the H$\beta$ corrected 
line fluxes and their associated errors for the ultraviolet,
optical and infrared spectra which we used.  

The flux calibration of the data is critical since we are comparing
optical, ultraviolet and infrared line fluxes.  The IUE
absolute fluxes have an error of about 10\% 
but the absolute fluxes of the optical and infrared spectra were not
as certain and depend on the photometric conditions under which the data
were obtained.  S91 showed that the absolute H$\beta$ line flux,
f(H$\beta$), can be fit by the equation
\begin{equation}
{\mbox{log(f(H}}\beta{\mbox{)) = -1.67 - 3.84log(t),}}
\end{equation}
\noindent where t is the number of days after S91's outburst date 
of 29 July 1984 (JD 2,445,911).  Figure 1 shows 
the S91 (squares) and KW (triangles) absolute H$\beta$ line fluxes
as a function of time and the least-squares fit to these fluxes.   
A91 reported f(H$\beta$) = 3.5 $\times$ 10$^{-12}$ 
ergs cm$^{-2}$ s$^{-1}$ on 4 July 1985 (t = 341).  
This value is {\it about 15\% smaller} than that
obtained from equation 1 for day 341.  A91's H$\beta$ line
flux is shown as a diamond in Figure 1.  For our analysis, we used the 
absolute H$\beta$ flux obtained from 
equation 1 for the date we chose to analyze.
The IUE spectrum (t=340), A91's optical spectra (t=341), the infrared 
lines (t=327) and the Mg\,{\sc ii} lines (t=330 and t=352)
are scaled, using equation 1, to the absolute H$\beta$ flux on their 
respective dates of observation.

The reddening to PW Vul has been obtained by various methods.
S91 estimated E(B-V) = 0.6 $\pm$ 0.06 from the average ratio of the
He\,{\sc ii} 1640 {\AA}/4686 {\AA} lines up to 2 years
after outburst.  A91 found that using E(B-V) = 0.55 $\pm$ 0.10 
removed the broad interstellar absorption feature
at 2175 {\AA} from IUE spectra.  
Williams et al. (1996) compared their Pa$\beta$
and Pa$\gamma$ line strengths of 20 March 1985 with KW's 
line strengths of H$\beta$ and H$\gamma$ on the same day.
They found A$_V$ = 1.78 $\pm$ 0.05, in excellent agreement with the
other reddening values.  In our analysis, we assumed an
E(B-V) = 0.55 $\pm$ 0.05 and used the extinction curve of Savage
and Mathis (1979).

\section{Optimization of photoionization model fits}

We used the photoionization code CLOUDY 90.02 (Ferland 1996) to 
predict the emission line fluxes relative to H$\beta$.  
CLOUDY simultaneously solves the equations of statistical and 
thermal equilibrium, incorporates collisional effects, and 
self-consistently treats energy balance and ionization.

CLOUDY requires numerous input 
parameters including the luminosity and the shape of the continuum
of the central source, the inner and outer shell radius, the hydrogen density, 
the filling factor, and elemental abundances.  In addition, CLOUDY allows 
the hydrogen density, n(r), and filling factor, f(r), to vary with the
radius as, 
\begin{equation}
{\mbox{n(r) = n(r$_{o}$)(r/r$_{o}$)$^{\alpha}$ cm$^{-3}$ and}}
\end{equation}
\begin{equation}
{\mbox{f(r) = f(r$_{o}$)(r/r$_{o}$)$^{\beta}$,}}
\end{equation}
\noindent where r$_{o}$ is the inner radius, and $\alpha$ and $\beta$ are
the exponents of the power law. 

Recently, powerful optimization techniques have been used 
to search multi-dimensional parameter space to find the best 
agreement between the predicted and observed line fluxes. 
Austin et al. (1996) used a Metropolis algorithm to determine the
abundances of V1974 Cyg 1992 and QU Vul 1984.  
Vanlandingham et al. (1996) and 
Schwarz et al. (1997) applied a gradient mapping algorithm 
from a minimization routine, MINUIT (James and Roos, 1993), 
to V838 Her 1991 and LMC 1991, respectively.  
The MINUIT routine uses a Davidson-Fletcher-Powell (DFP) variable metric
algorithm to search for a minimum in parameter 
space (see Press et al. 1992 for details).

For the abundance determinations of PW Vul, we used MINUIT
as a driver for CLOUDY.  Following an initial guess of the input
parameters, CLOUDY calculated the ratios of the emission line 
fluxes,$M_i$, to H$\beta$.  
The goodness of fit was estimated from the $\chi^2$ of 
the model:
\begin{equation}
\chi^2 = \sum_{i} \frac{(M_{i} - O_{i})^2}{(\sigma_{i})^2},
\end{equation}
\noindent where $O_i$ is the observed line flux ratio to H$\beta$
and $\sigma_{i}$ is the error associated with the observed line.  
The number of degrees of
freedom in the model, $\nu$, is equal to the number of 
lines modeled minus the number of parameters used in the model.
A good model has a $\chi^2 \approx \nu$.  

In Figure 2 we show the lowest $\chi^2$ fit determined by MINUIT
to the early July observations and the relative
contribution of each line to the overall $\chi^2$.
The model had 14 parameters, $\nu$ = 12 and a $\chi^2$ of 12.  
The best-fitting model had a blackbody continuum with an 
effective temperature for the central source of 2.4 $\times$ 10$^5$ K.  
The progression from the early, cool ($\approx$ 1-2 $\times$ 10$^4$ K),
optically thick atmosphere (Schwarz et al. 1997) to an
optically thin, nebular shell illuminated by a hot source
is consistent with the rise in X-rays seen by EXOSAT from
100 to 300 days after outburst (\"{O}gelman et al. 1987).  
The luminosity was 6.3 $\times$ 10$^{37}$ ergs s$^{-1}$, which is below 
the Eddington luminosity for a one solar mass white dwarf. 
The hydrogen number density of the best model was 
1.1 $\times$ 10$^7$ cm$^{-3}$ at the inner radius and decreased 
as a power law with an exponent of $\alpha$ = -0.3.
 
S91 obtained expansion velocities of 600 km s$^{-1}$ from the width 
of H$\beta$, consistent with KW's FWHM measurements of H$\alpha$, and 
400 km s$^{-1}$ from double peaks observed in [O\,{\sc iii}] (5007 {\AA})
in nebular spectra obtained one to two years after outburst.
Ringwald $\&$ Naylor (1996) measured an H$\alpha$ expansion 
velocity of about 500 km s$^{-1}$ seven years after outburst.
Observations taken very early in the outburst (Andrillat 1984, Rosino et al.
1984) found emission line widths consistent with expansion velocities 
of $\approx$ 1200 km s$^{-1}$.  The inner radius of our model,
1.8 $\times$ 10$^{15}$ cm (equal to a constant expansion velocity
of 610 km s$^{-1}$ for 340 days), was consistent with the lower
expansion velocities reported.  The outer radius was 
determined to be 7.6 $\times$ 10$^{15}$ cm.  

Recent HST images of V1974 Cyg 1992 (Paresce et al. 1995) and high
resolution spectra of other novae indicate 
that the ejecta have a clumpy structure (Shore et al. 1993).  
Our analysis predicted a low filling factor that varies with radius,
consistent with the structure seen in the emission lines.
The filling factor of the best model was 0.025 at the inner radius and
decreased as a power law with an exponent of $\beta$ = -0.6.
The hydrogen mass contained in this model shell was calculated from,
\begin{equation}
{\mbox{M$_{shell}$ = n(r$_{o}$) f(r$_{o}$)}} \int_{R_{inner}}^{R_{outer}} 
{\mbox{(r/r$_{o}$)$^{\alpha+\beta}$ dV(r)},}
\end{equation}
\noindent where V(r) is the volume of the model.  Using the model 
parameters, a hydrogen shell mass of 1.6$\times$ 10$^{-4}$ M$_{\odot}$ was 
determined with a range of uncertainty between  3 $\times$ 10$^{-3}$
and 3$\times$ 10$^{-6}$ M$_{\odot}$.  Our derived ejected mass was 3 times
the value that Stickland et al. (1981) derived for the fast 
CO nova V1668 Cyg 1978.  This high an ejected mass is consistent 
with the slow speed class of PW Vul. Slow novae are believed to occur on
less massive white dwarfs than fast novae and can, therefore, accrete
more mass before the explosion.

The helium, carbon, nitrogen, and oxygen abundances all showed enhanced
values relative to solar with nitrogen being the highest at 85 times solar.
The magnesium and iron abundances were found to be one and two times 
the solar values, respectively.  These abundances are estimates only
since the Mg\,{\sc ii} 2800 {\AA} doublet and 
the [Fe\,{\sc vii}] 6087 {\AA} line are the only features of those species
used in the analysis and [Fe\,{\sc vii}] 6087 {\AA} is
quite weak. No neon lines are reported by A91
but a solar neon abundance predicted a Ne\,{\sc iii} (3869 and 3967 {\AA})
line flux that was consistent with earlier values 
reported by KW on 2 June 1985.  In Figure 3, we compare our derived 
abundances with those of S91 and A94.

In Table 2, we present the parameters and 
uncertainties from our lowest $\chi^2$ model for early July.
To estimate the uncertainty in the parameters, the model
was recomputed, increasing or decreasing one parameter
at a time.  The amount by which the parameter could
be changed and still provide an acceptable $\chi^2$ corresponds to the
uncertainty.  $\chi^2 \approx$ 23 was 
deemed the lowest $\chi^2$ that a model could have and still provide an 
acceptable fit to the data. 

We then increased E(B-V) to 0.60, in order to estimate 
how the uncertainty in the reddening affected the derived abundances.  
The larger reddening mainly affects the higher ionization lines in 
the ultraviolet.  
The best model now required a combination of about twice the
carbon, nitrogen, oxygen and magnesium abundances as before, a 
lower luminosity, (2.2 $\times$ 10$^{37}$ ergs s$^{-1}$), 
and smaller inner and outer radii 
(1.4 and 4 $\times$ 10$^{15}$ cm).  The increase in the
abundances increased all line ratios so as to fit the higher reddening but 
overestimated the strongest lines in the ultraviolet.  The decrease
in the luminosity compensated for this effect by decreasing the
high ionization emission line strengths relative to the low ionization lines.

As a check of our abundance solution, we modeled KW's optical spectrum 
of 25 April 1986 and the IUE spectra SWP28068 and LWP7925 on 31 March 1986.
The line fluxes were scaled to the absolute H$\beta$ flux on their 
respective dates with equation 1.  Since only line fluxes of H$\beta$ 
and the strong [O\,{\sc iii}] lines (4363, 4959 $\&$ 5007 {\AA}) were 
reported by KW, the total number of line flux ratios available to be modeled
was only 13.  The model had an equal number of parameters and data points
making it impossible to obtain a statistically reasonable fit for this date. 
However, we can use the best fit model as a confirmation of 
our previous solution.  As a starting point for MINUIT, we used the 
same luminosity, blackbody temperature and inner filling factor 
used for July 1985.  We assumed a lower constant hydrogen density
($\alpha$ = 0) and an initial inner radius consistent 
with a constant expansion 
velocity of 610 km s$^{-1}$ over 670 days.  The best fit 
($\chi^2$=4) was found with an effective temperature of 2.2 $\times$ 10$^{5}$
K, a constant hydrogen density of 10$^6$ cm$^{-3}$,  a 
constant filling factor of 0.016 and an inner radius of
4 $\times$ 10$^{15}$ cm.  The outer radius was 3.2 $\times$ 10$^{16}$ cm.
The ejected mass of this model, 2 $\times$ 10$^{-3}$ M$_{\odot}$, is over a 
factor of 10 higher than before and confirms the high ejected mass
derived in July 1985.  Our fit indicated that the luminosity and 
blackbody temperature did not change much in the ten months between
the dates modeled.  
The abundances of helium, carbon, nitrogen, oxygen and magnesium
were 1, 5, 72, 6.8, and 0.8 times solar abundances, respectively.
These abundances are well within the errors of previously calculated 
abundances for July 1985 and support our abundance determinations for PW Vul.

\section{Abundance Determinations of Saizar et al. (1991)}

To derive the elemental abundances of helium, nitrogen, and
oxygen.  S91 used 4 to 6 optical spectra taken from 
early 1985 until mid 1987.  The carbon abundance was determined
using two ultraviolet spectra that were contemporaneous with optical
spectra.  The final elemental abundances were averages of all 
the observations in which the elements were observed. 

S91 obtained the helium to hydrogen abundance from:

\begin{equation}
\frac{He}{H} \approx \frac{HeII}{HII} + \frac{HeIII}{HII}.
\end{equation}

\noindent The ion abundances were given by: 

\begin{equation}
\frac{HeII}{HII} = \frac{j(HeI)}{j(H\beta)}
\frac{\alpha_{H\beta}}{\alpha_{HeI}}\frac{5876}{4861}
\end{equation}
\begin{equation}
\frac{HeIII}{HII} = \frac{j(HeII)}{j(H\beta)}
\frac{\alpha_{H\beta}}{\alpha_{HeII}}\frac{4686}{4861}
\end{equation}

\noindent where $\alpha$ is the recombination coefficient and j/j(H$\beta$)
is the line ratio to H$\beta$.  S91 accounted for collisional excitation from
the metastable 2s $^3$S level of He\,{\sc i} by using the correction of
Clegg (1987).  The HeII/HII ratio was multiplied
by (1+C/R)$^{-1}$ where C/R is the ratio of excitations by collisions to
recombinations and is tabulated by Clegg.

To derive the metal abundances, S91 assumed that the ratio of 
elemental abundances was approximately the same as the ratio of 
ion abundances for elements with similar ionization potentials.  
S91 used the following approximations to determine the elemental 
ratios from optical spectra:

\begin{equation}
\frac{O}{H} \approx \frac{OIII}{HeII}\Bigr(\frac{HeII + HeIII}{HII}\Bigr) = 
\frac{OIII}{HII}\Bigl(1 + \frac{HeIII}{HII}\frac{HII}{HeII}\Bigr)
\end{equation}
\begin{equation}
\frac{N}{H} 
\approx \frac{NII}{OII}\Bigr(\frac{OII + OIII}{HII}\Bigr) =
\frac{NII}{HII}\Bigl(1 + \frac{OIII}{HII}\frac{HII}{OII}\Bigr)
\end{equation}

\noindent The ion abundances relative to hydrogen are given by:

\begin{equation}
\frac{OIII}{HII} = 
\frac{j(\lambda\lambda4959+5007)}{j(H\beta)}\frac{5007}{4861}
\Bigl(\frac{N_{e} \alpha_{H\beta}}{A_{^{1}D_{2}-^{3}P} N_{^{1}D_{2}}}\Bigr)
\end{equation}
\begin{equation}
\frac{NII}{HII} = 
\frac{j(\lambda5755)}{j(H\beta)}\frac{5755}{4861}
\Bigl(\frac{N_{e} \alpha_{H\beta}}{A_{^{1}S_{0}-^{1}D_{2}} N_{^{1}S_{0}}}\Bigr)
\end{equation}
\begin{equation}
\frac{OII}{HII} = 
\frac{j(\lambda\lambda7320+7330)}{j(H\beta)}\frac{7325}{4861}
\Bigl(\frac{N_{e} \alpha_{H\beta}}
{N_{^{2}P_{1/2}}(A_{^{2}P_{1/2}-^{2}D_{3/2}}+A_{^{2}P_{1/2}-^{2}D_{5/2}})+
N_{^{2}P_{3/2}}(A_{^{2}P_{3/2}-^{2}D_{3/2}}+A_{^{2}P_{3/2}-^{2}D_{5/2}})}\Bigr)
\end{equation}

\noindent where N$_e$ is the electron number density, A is the 
transition probability, and N is the relative fraction 
of ions in the given level.  Using the ultraviolet spectra, the carbon
to oxygen abundance was found from,

\begin{equation}
\frac{C}{O} \approx \frac{CIII}{OIII} = 
\frac{j(\lambda\lambda1907+1909)}{j(\lambda\lambda1663+1666)}
\frac{1908}{1665}\Bigl(
\frac{A_{^{5}S-^{3}P} N_{^{5}S}(OIII)}{A_{^{3}P-^{1}S} N_{^{3}P}(CIII)}\Bigr).
\end{equation}

Since S91 never used the ratio of ultraviolet to optical lines, their 
analysis was not affected by errors in the absolute flux calibration
of the optical spectra.  The uncertainty in the
reddening was also less important since the abundances were determined 
from line ratios of similar wavelengths.  In addition,
S91's analysis had the advantage of multiple observations over 
two years. 

From the [O\,{\sc iii}] line ratios 
j($\lambda\lambda$1660+1666)/j($\lambda\lambda$4959+5007) and
j($\lambda$4363)/j($\lambda\lambda$4959+5007), S91 determined the 
electron temperatures and densities for the four dates that both
optical and ultraviolet spectra existed.  S91 found a constant
electron temperature of 13,200 K and an electron density decreasing
from 7.6 $\times$ 10$^6$ cm$^{-3}$ on day 253 to 1.7 $\times$ 10$^5$
cm$^{-3}$ on day 681.  

The derived electron temperature and density 
were assumed to be constant throughout the shell for each date.  This
assumption gives at best only a rough abundance solution since higher
ionization ions are expected to form in higher temperature and 
denser regions than the low ionization ions.  
This can be seen in Figure 4 where the
electron density, electron temperature, and relative ion
concentrations of our best July model are presented as a function of 
radius.  To determine the effect of using more realistic electron 
temperatures and densities for each different region of the shell
on the final elemental abundances, we used the average electron
temperature and density (from the region where the ion 
was dominant in our best July model) 
to obtain the ion abundances relative to ionized 
hydrogen in equations 6-14.  The average electron temperatures 
and densities of the various ion regions are given in Table 3.
We used the line ratios reported in S91 and the atomic constants 
of Mendoza (1983).  The relative level populations were calculated 
with a three level atom for [O\,{\sc iii}] and [{N\,{\sc ii}],
a five level atom for [O\,{\sc ii}], and a simple two level atom
for the C\,{\sc iii}]/O\,{\sc iii}] calculation.  In Table 4, we 
compare the elemental abundances of S91 with the abundances derived 
from our model averaged electron temperatures and densities. 
The model averaged electron temperatures and densities are
more realistic and, in conjunction with S91's measured line fluxes, give 
higher abundances that are much closer to the abundances of our best 
fitting July photoionization model.  The primary reason for the increase seen 
in these abundances is the generally lower electron temperatures.
The lower electron temperatures produce fewer ions 
in excited levels and thus the total ion abundances 
must be increased to produce the same observed emission line strengths.

\section{Abundance Determinations of Andre\"{a} et al. (1991,1994)}

When modeling combined
ultraviolet and optical spectra, an accurate flux calibration is 
needed between the spectral regions.  To see why, consider that 
during the nebular phase the ultraviolet contains strong emission lines of
high ionization potential (N\,{\sc v} 1240 {\AA}, N\,{\sc iv}] 1486 {\AA},
N\,{\sc iii}] 1750 {\AA}, C\,{\sc iv} 1550 {\AA}, C\,{\sc iii}] 1909 {\AA},
O\,{\sc iv}] 1401 {\AA}, and O\,{\sc iii}] 1666 {\AA}) while the optical
contains strong emission lines from species with a lower ionization potential
([N\,{\sc ii}] 5755 {\AA}, [O\,{\sc iii}] 4959,5007 {\AA}, [O\,{\sc ii}]
7325 {\AA}, and [O\,{\sc i}] 6300, 6363 {\AA}).  If the absolute optical
calibration is underestimated so that
the optical fluxes are too low, then the ratio of
ultraviolet/optical lines will be too large and the ionic abundances of the
lower ionization potential ions will be too low.  An analysis of the
line fluxes will lead to preferentially populated higher ion stages.
This creates the false impression that a substantial fraction
of the elements are in higher (unobserved) ionization stages,
leading to over estimates of the elemental abundances.

As stated in \S2, A91 obtained an H$\beta$ flux that was about
15$\%$ smaller than that given by equation 1.  We will address how
this affected the abundance analysis later in this section 
but for now we use only the line fluxes as measured by A91.  
Adopting the values of A91 allow us to
directly compare the abundances determined by A91 and A94 with our
best fits to their data using our techniques.

A91 applied ionization correction factors (ICFs) to estimate 
the fraction of an element in unobserved ionization stages.
This method assumes that an ICF derived for one 
element holds for {\it all} other elements and is also
dependent on the number of the ionization stages observed.
A91 used simultaneous optical and UV spectral observations to 
obtain as many elemental ionization stages as possible.  

The ICF for each element was determined by A91 from:

\begin{equation}
ICF = \frac{C}{(I_{2} - I_{1})}
\end{equation}

\noindent where $C$ is a constant and $I_2$ and $I_1$ are the ionization
potentials of the highest and lowest ion observed, respectively.
A91 estimated that 95$\%$ (ICF = 1.05) of the
nitrogen was observed in the ionization stages N\,{\sc ii} to N\,{\sc vi}
giving C = 87.6 eV.  This constant, derived solely from the nitrogen
ion abundances, was used in equation 18 to determine the ICF for all
other elements.

As a check of the accuracy of A91's ICFs, we used the same parameters
and abundances used by A94 (A94's photoionization model $\#$1) in
CLOUDY to obtain the volume averaged ionization fractions.  
{\it We note that we are only comparing the ICFs determined by 
A91 and not the methods by which the ion abundances were derived.}
In Table 5, we present the ion abundances
as a fraction of the elemental abundance for the first six ionization
stages.  The last column gives the ICFs determined for unobserved ions.  
The more detailed ionization calculations of CLOUDY 
predict more neutral nitrogen and oxygen and less 
N\,{\sc iv} and O\,{\sc vi} than A91 but the
ICFs are in close agreement with those determined by A91.  The
agreement in ICFs is not surprising, since A91 observed both nitrogen and
oxygen in five ionization states and estimated that 
they had accounted for a majority of the abundances of
these elements.  The differences between 
ICFs and photoionization modeling were more clearly shown in carbon,
where only three ionization states were observed.  The carbon ICF of A91
was significantly larger then the ICF of CLOUDY and thus would predict
1.4 times more carbon than CLOUDY.  As expected, the ICF method 
provides poor abundance solutions when too few 
ionization stages of an element are available in the spectra. 

The abundances derived by A91 were checked in A94 by comparing
with the results from a photoionization model.  A94 reported that their
best fit model had slightly higher abundances of nitrogen, oxygen and iron 
and 20$\%$ less carbon than found by A91 from the ICF method.  
We assumed this to mean a 5$\%$ increase in nitrogen, oxygen and iron  
from their ICF method abundances. The difference in carbon abundance 
between the two methods again reflects an inaccuracy 
in the ICF method when insufficient ionization stages are observed.

We then used A94's best photoionization model parameters (their model $\#$2) 
in CLOUDY to check the consistency of the two codes.   
In Table 6 we present the emission line strengths predicted 
by CLOUDY and A94, A94's measured emission line strengths, and
the average line errors given by A91.  The biggest difference is
seen in the predictions of C\,{\sc iv} 1550 {\AA} and C\,{\sc iii}]
1909 {\AA}.  All other lines are within 20-30$\%$ of one another.  

Clearly there are differences between the two photoionization codes.  To
determine if the differences are significant, MINUIT optimized the
A94 parameters with respect to the measured line fluxes to find the
best fit parameters.  Ideally, the minimized parameters should
not be too different from their best model.  The
parameters given by A94 were used as the starting points for MINUIT. 
We did this twice, first with the filling factor and the radial
dependence of the density and filling factor allowed to vary.  The 
second had the filling factor set equal to 1 and the 
radial dependence of the density and filling factor set to zero
to duplicate the assumptions of A94.  The results are 
given in Table 7 along with the $\chi^2$ of the two models.  
The physical parameters for both minimized models were
close to those of A94.  When the filling factor was not fixed, MINUIT
found the lowest $\chi^2$ for a low constant
filling factor of 0.045, consistent with the observations and our results.  
The CNO abundances from the low filling factor 
model were greater than those obtained in the fixed filling factor model.

In our models, we generally find that if the filling factor is 
decreased, then the luminosity, density and abundances have to be increased 
(with all other parameters roughly the same) in order to fit the 
same measured line fluxes.  A decrease in the filling factor causes
higher predicted line fluxes in [O\,{\sc i}] (6300, 6363 {\AA}), 
[O\,{\sc II}] (7320, 7330 {\AA}) and [N\,{\sc ii}] (6548, 6584 {\AA}).
In the lower filling factor model, these lines form in a region with a 
larger electron density than the 
same lines in the high filling factor model.
The decrease in the filling factor allows the more energetic
photons to penetrate further into the shell and the electron 
density increases as a result of the increased ionization.  
This leads to an increase in the number of 
collisional excitations and, in turn, to
the increased line emission observed.  The situation can be somewhat 
alleviated by increasing the luminosity and density but the best model 
requires slightly higher abundances to match the measured line fluxes.

The fixed filling factor model used all the parameters included
in A94's model and thus is the model to compare to A94's model
for a predictive test of the photoionization codes.  The abundances 
were within 30$\%$ of A94's values except for carbon. 
The carbon abundance is less than
half that found by A94, indicating a factor of $\geq$2 disagreement 
between the models.  Even the best fit model with a more realistic
filling factor did not provide enough carbon to reproduce 
A94's value.  A detailed study of the reasons for the
carbon differences is beyond the scope of this paper.  We 
conclude that the choice of the photoionization code was responsible for
the major contribution to the differences in the carbon abundance.  

To estimate the effect of a difference in the flux
calibration on the abundances,
we found the best fit to our July 1985 data using the H$\beta$
flux given by A91 instead of equation 1.  The primary result of 
not including this correction was higher ultraviolet line fluxes 
relative to H$\beta$, resulting in higher carbon, nitrogen, 
oxygen and magnesium abundances.  We started our minimization process with the 
same parameters as in our H$\beta$ flux corrected model 
but with higher CNO abundances. The best fit model had roughly the same
physical parameters except it is 1/3 as luminous and has 
1/3 the inner radius as the flux corrected model.  In addition,
the best fit model had 1.5, 1.6 and 2 times more C, N and O elements  
than the flux corrected model.

\section{Conclusion}

We have determined our own set of elemental abundances
for the slow nova PW Vul 1984 from the largest set of available
contemporaneous UV, optical and infrared spectra.  The abundances
were found using a photoionization code, combined with a minimization
technique to search parameter space for the best agreement between
the predicted and observed emission line fluxes.   From our 
lowest $\chi^2$ model, we found that helium, carbon, nitrogen, 
and oxygen showed enhancements of
1.0 $\pm$ 0.4, 7.0 $^{+7}_{-4}$, 85 $^{+59}_{-41}$ and 6 $^{+7}_{-2}$ 
times the solar value, respectively.  In addition, we 
found weak evidence (modeling only 1 weak line each) that neon and 
magnesium had solar abundances while Fe had twice the solar abundance. 
The expansion velocity derived from the inner radius, $\approx$ 600 km
s$^{-1}$, and the blackbody effective temperature,
2.4 $^{+0.27}_{-0.4} \times$ 10$^5$ K of our photoionization 
model are consistent with X-ray and high resolution optical spectra.
The low value of the filling factor is consistent with both the 
observed line profiles and the clumpiness seen in HST images of 
the ejecta of nova V1974 Cygni 1992 and other novae.
We determined that our uncertainty in the reddening equated to a 
factor of 2 in the abundances.  We applied the same techniques
to optical and UV spectra obtained 10 months after our best fit 
as a predictive test of our elemental abundances and parameters. 
The blackbody temperature and luminosity
were approximately the same and the radii were consistent with
the expansion determined from the previous best fit model.  The hydrogen
density and filling factor decreased and all of the abundances were 
well within the error range established in the first model, confirming
our abundance solution.
 
Theoretical modeling predicts that the nova outburst occurs as the
result of thermonuclear runaways in the accreted hydrogen rich
envelopes of white dwarfs (Starrfield 1989).  The abundance analyses of 
nova ejecta show, for most novae, that there has been
significant mixing of the accreted material with core material which 
has either a CO or an ONeMg composition (Starrfield et al. 1997). This
enrichment appears as large overabundances of a variety of
elements from carbon through silicon depending on the particular
nova. In the case of PW Vul, we find that the CNO elements comprise 
about 13\% of the mass of the ejecta.  In contrast, the 
CNO elements comprise 1.3\% of solar material.
The amount and distribution of the enrichments
implies mixing with a CO white dwarf core.  The abundances of Ne and 
Mg are solar which supports our contention that the core material does not
contain enriched neon and magnesium.  Finally, our model gives an
ejected mass of $\sim 1.6 \times 10^{-4}$M$_{\odot}$  
which suggests that the outburst took place on a white dwarf 
with a mass of $\sim 1.05$M$_{\odot}$ (Starrfield 1989).  

We studied the methods used by S91, A91 and A94 and were able to
reproduce the abundances reported by these authors using their 
data and methods.  
S91 used a plasma diagnostic approach to obtain
the average ion abundances from spectra obtained between 
1 and 3 years after outburst (primarily optical).
Ratios of ions with similar ionization potentials were assumed
to be equal to their elemental abundance ratios.  Applying these
assumptions but using more realistic electron temperatures 
and densities gave higher abundances which were 
similar to those determined from our best photoionization
model.   We conclude that the abundances of S91 are 
significantly lower then ours because of the higher electron 
temperatures used by S91.

A91 used the ICF method to obtain their abundances.  This 
technique assumed a correction to the elemental abundances caused by
unobserved ionization states.  Contemporaneous
optical and ultraviolet spectra were used to maximize the 
number of observable ionization states of as many elements as possible.  
A94 then used a photoionization code to verify the results of 
A91's ICF method.  We used the parameters and line flux measurements
given by A94 
in our photoionization code to assess the accuracy of the ICF method.  For 
elements that had many observed ionization stages such as nitrogen 
and oxygen, we found that the ICF method is a good approximation 
but when fewer ionization stages were observed, such as for carbon,
the approximation was poor.  For example, the ICF method 
predicted 1.4 times more carbon than was present in our 
model and 20$\%$ more than in A94's model.

A consistency check between CLOUDY and the photoionization model used
by A94 showed persistent differences in the carbon abundances.  We optimized
A94's best parameters to the measured emission line strengths of A91 to see
how close the final best fit parameters compared with A94's values.
All of our parameters were within 20-30$\%$ of A94 except for
carbon which had only 1/2 of A94's abundances.  These results imply
that the two codes produce consistent results under these 
circumstances except for a factor of 2 in the derived carbon abundance.

The ICF and photoionization modeling of optical and ultraviolet data
required that the spectra be absolutely flux calibrated.  A91 found
15\% less H$\beta$ flux than calculated by S91 from 
observations of the decline
of the H$\beta$ flux over 2 years of observations.  Modeling our data with
the H$\beta$ absolute flux of A91 produced carbon, nitrogen, oxygen,
and magnesium abundances that were over 2 times greater then our 
best fit models with our value of H$\beta$.
{\it This underscores the need for contemporaneous
and absolute flux calibrated optical and UV spectral observations 
to accurately determine the elemental abundances.}
A91's analysis consisted of simultaneous optical and UV
spectra at only {\it one} epoch and thus their results 
depended on the value of the flux of H$\beta$.  Increasing
our best fit abundances by a factor of 2 (to roughly compensate for
the H$\beta$ flux used by A91) and increasing the carbon abundance
by another factor of 2 (to compensate for the differences in
the photoionization codes) gave high elemental abundances 
roughly equal to those reported by A94.

\section{Acknowledgments}

It is a pleasure to thank G. Ferland for use of his photoionization
code CLOUDY and J. Truran, H. Drechsel and P. Saizar for stimulating 
discussions.  This work was supported in part by NASA and NSF grants 
to Arizona State University.

\clearpage 
\section{Tables}
\begin{table}
\caption{Line fluxes relative to H$\beta$ from equ. 1}
\label{symbols}
\begin{tabular}{@{}lll}
Element ({\AA}) & flux/flux(H$\beta$) & error ($\%$) \\
N\,{\sc v} 1240 & 1.1 & 30$^a$ \\
C\,{\sc ii} 1335 & 0.12 & 45$^a$ \\
O\,{\sc iv} ] 1402 & 0.27 & 35$^a$ \\
N\,{\sc iv} ] 1486 & 1.7 & 30$^a$ \\
C\,{\sc iv} 1550 & 2.6 & 30$^a$ \\
He\,{\sc ii} 1640 & 0.29 & 35$^a$ \\
O\,{\sc iii}] 1666 & 0.24 & 35$^a$ \\
N\,{\sc iv} 1718 & 0.06 & 65$^a$ \\
N\,{\sc iii}] 1750 & 1.2 & 30$^a$ \\
C\,{\sc iii}] 1909 & 1.4 & 30$^a$ \\
Mg\,{\sc ii} 2800 & 0.42 & 40$^b$ \\
H$\gamma$ 4340 & 0.37 & 40$^c$ \\
{[O\,{\sc iii}]} 4363 & 2.1 & 40$^c$ \\
He\,{\sc ii} 4686 & 0.34 & 30$^c$ \\
H$\beta$ 4861 & 1.00 & 15$^c$ \\
{[O\,{\sc iii}]} 4959 & 2.7 & 30$^c$ \\
{[O\,{\sc iii}]} 5007 & 7.9 & 30$^c$ \\
{[N\,{\sc ii}]} 5755 & 1.6 & 30$^c$ \\
He\,{\sc i} 5876 & 0.14 & 30$^c$ \\
Fe\,{\sc v ii}] 6087 & 0.21 & 30$^c$ \\
{[O\,{\sc i}]} 6300 & 0.17 & 50$^c$ \\
{[O\,{\sc i}]} 6363 & 0.11 & 30$^c$ \\
{[N\,{\sc ii}]} 6548 & 0.37 & 65$^c$ \\
H$\alpha$ 6563 & 4.4 & 30$^c$ \\
{[N\,{\sc ii}]} 6584 & 1.1 & 65$^c$ \\
He\,{\sc i} 10830 & 2.9 & 35$^d$ \\
Pa$\gamma$ 10938 & 0.19 & 45$^d$ \\
Pa$\beta$ 12818 & 0.47 & 35$^d$ \\
\end{tabular}

\medskip
{\em $^a$} UV: SWP26342 on 3.09 July 1985.

{\em $^b$} UV: average of LWP6264 on 24.6 June 1985 and LWP6408
on 16.7 July 1985.

{\em $^c$} Optical: Andre\"{a} et al. (1991) on 1-6 July 1985.

{\em $^d$} Infrared: Williams et al. (1996) on 21 June 1985.

\end{table}

\clearpage
\begin{table}
\caption{Parameters of best fitting model for June/July observations}
\label{symbols}
\begin{tabular}{@{}ll}
Parameter & Value \\
Black body temperature & 2.4 $^{+0.27}_{-0.40}$ $\times$ 10$^5$ K \\
Source luminosity & 6.3 $^{+1.6}_{-3.3}$ $\times$ 10$^{37}$ ergs s$^{-1}$ \\
Hydrogen density & 1.1 $^{+1.4}_{-0.1}$ $\times$ 10$^7$ cm$^{-3}$ \\
$\alpha^{a}$ & -0.3 $^{+0.3}_{-0.13}$ \\
Inner radius & 1.8 $^{+1.9}_{-1.3}$ $\times$ 10$^{15}$ cm \\
Outer radius & 7.6 $\times$ 10$^{15}$ cm$^{b}$ \\
Filling factor & 0.025 $^{+0.055}_{-0.013}$ \\
$\beta^{c}$ & -0.6 $^{+0.6}_{-0.9}$ \\
He/He$_{\odot}^d$ & 1.0 $\pm$ 0.4 \\
C/C$_{\odot}^d$ & 7.0 $^{+7}_{-4}$ \\
N/N$_{\odot}^d$ & 85 $^{+59}_{-41}$ \\
O/O$_{\odot}^d$ & 6 $^{+7}_{-2}$ \\
\end{tabular}

\medskip

{\em $^a$} Radial dependence of the density r$^{\alpha}$.

{\em $^b$} Model is insensitive to an upper limit for the outer radius 

because it is radiation bounded.

{\em $^c$} Radial dependence of the filling factor r$^{\beta}$.

{\em $^d$} Log of the solar number abundances relative to hydrogen 

He:-1.0 C: -3.45 N:-4.03 O: -3.13 (Grevesse $\&$ Noel 1993).  Estimates

on the Ne, Mg and Fe abundances are -3.93, -4.42 and -4.2.

\end{table}

\clearpage
\begin{table}
\caption{Averaged electron temperatures and densities of best fit model}
\label{symbols}
\begin{tabular}{@{}lrl}
Element & T$_{e}$ (K) & N$_e$ (cm$^{-3}$) \\
He III   & 15,000 & 1.3 $\times$ 10$^7$ \\
H II     & 14,000 & 1.1 $\times$ 10$^7$ \\
He II    & 11,000 & 8.5 $\times$ 10$^6$ \\
C III    & 11,000 & 8.8 $\times$ 10$^6$ \\
O III    & 11,000 & 8.8 $\times$ 10$^6$ \\
N II     &  9,000 & 6.0 $\times$ 10$^6$ \\ 
O II     &  9,000 & 6.0 $\times$ 10$^6$ \\
\end{tabular}
\end{table}

\clearpage
\begin{table}
\caption{Abundance comparison to S91}
\label{symbols}
\begin{tabular}{@{}lrr}
Element$^a$ & S91 & CLOUDY$^b$ \\
He/He$_{\odot}$ & 0.9 & 1.0 \\
C/C$_{\odot}$ & 1.1 & 2.8 \\
N/N$_{\odot}$ & 55 & 139 \\
O/O$_{\odot}$ & 1.8 & 7.3 \\
\end{tabular}

\medskip
{\em $^a$} Log of the solar number abundances relative to hydrogen 

He:-1.0 C: -3.45 N:-4.03 O: -3.13 (Grevesse $\&$ Noel 1993).

{\em $^b$} Abundances from our electron temperatures and densities in equ 12-16.

\end{table}

\clearpage
\begin{table}
\caption{Comparison of ICF derived by A91 and predicted by CLOUDY}
\label{symbols}
\begin{tabular}{@{}lllllllll}
Element & Method & I  & II  & III & IV  & V  & VI  &  ICF \\
Carbon & A91 & nl & 0.18 & 0.32 & 0.10 & nl & nl & 1.65 \\
Carbon & CLDY & 0.0 & 0.31 & 0.41 & 0.10 & 0.18 & 0.00 & 1.22 \\
  &  &  &  &  &  &  &  & \\
Nitrogen & A91 & 0.00 & 0.08 & 0.45 & 0.35 & 0.08 & nl & 1.05 \\
Nitrogen & CLDY & 0.22 & 0.10 & 0.41 & 0.14 & 0.08 & 0.06 & 1.04 \\
  &  &  &  &  &  &  &  & \\
Oxygen  & A91 & 0.02 & 0.16 & 0.29 & 0.41 & nl & nl & 1.10 \\
Oxygen  & CLDY & 0.27 & 0.05 & 0.41 & 0.14 & 0.09 & 0.04 & 1.15 \\
\end{tabular}

\medskip
nl: line not observed in spectra. 

\end{table}

\clearpage
\begin{table}
\caption{Comparison of CLOUDY and A94's photoionization model}
\label{symbols}
\begin{tabular}{@{}llll}
Element ({\AA}) &  CLOUDY$^a$ & A94 model$^{a}$ & 
A94 measured$^{ab}$ \\
N\,{\sc v} 1240 &   36.5 & 28.8 & 30.4 $\pm$ 15$\%$ \\
C\,{\sc ii} 1335 &  3.7  &  3.7 & 2.0 $\pm$ 40$\%$ \\
{O\,{\sc iv} ]} 1402 &  2.6  &  3.3 & 4.0 $\pm$ 40$\%$ \\
{N\,{\sc iv} ]} 1486 &  24.5 & 24.0 & 20.9 $\pm$ 15$\%$ \\
C\,{\sc iv} 1550 &  46.9 & 27.3 & 30.0 $\pm$ 15$\%$ \\
He\,{\sc ii} 1640 & 2.7  &  2.1 &  2.4 $\pm$ 40$\%$ \\
{O\,{\sc iii}]} 1666 & 2.2  &  2.8 & 2.0 $\pm$ 40$\%$ \\
N\,{\sc iv} 1718 &  1.1  & 0.85 & 0.65 $\pm$ 75$\%$ \\
{N\,{\sc iii}]} 1750 & 10.6 & 10.0 & 12.6 $\pm$ 15$\%$ \\
{C\,{\sc iii}]} 1909 & 21.1 & 14.9 & 18.3 $\pm$ 15$\%$ \\
{[O\,{\sc ii}]} 2471 &  0.19 & 0.34 & 0.83 $\pm$ 75$\%$ \\
Mg\,{\sc ii} 2800 & 0.42 & 0.55 & 1.80 $\pm$ 75$\%$ \\
H$\gamma$ 4340 &    0.49 & 0.46 & 0.52 $\pm$ 40$\%$ \\
{[O\,{\sc iii}]} 4363 & 2.65  & 2.7  & 2.7 $\pm$ 15$\%$ \\
He\,{\sc ii} 4686 & 0.29 & 0.32 & 0.37 $\pm$ 15$\%$ \\
{[O\,{\sc iii}]} 4959 & 3.3  & 2.6  & 2.5 $\pm$ 15$\%$ \\
{[O\,{\sc iii}]} 5007 & 9.4  & 7.8  & 7.5 $\pm$ 15$\%$ \\ 
{[N\,{\sc ii}]} 5755 & 0.68  & 1.0 & 1.1 $\pm$ 15$\%$ \\
He\,{\sc i} 5876 & 0.11 & 0.13 & 0.10 $\pm$ 15$\%$ \\
H$\alpha$ 6563 & 2.8 & 2.9 & 2.3 $\pm$ 15$\%$ \\
\end{tabular}

\medskip
{\em $^a$} Line fluxes relative to H$\beta$. 

{\em $^b$} Error estimates from A91.

\end{table}

\clearpage
\begin{table}
\caption{Parameters of best photoionization model fit to A91 line fluxes}
\label{symbols}
\begin{tabular}{@{}llll}
Parameter & CLDY+MIN$^{a,c}$ & CLDY+MIN$^{b,c}$ &  A94 \\
BB temp ($\times$10$^5$ K) & 2.3 & 2.3 & 2.5 \\
Lum ($\times$10$^{38}$ ergs s$^{-1})$ & 1.9 & 1.3 & 1.6 \\
H Den ($\times$10$^7$ cm$^{-3}$) & 1.3 & 1.1 & 1 \\
Radius ($\times$10$^{15}$ cm) & 4 & 4.4 & 4.3 \\
$\alpha$ & -0.1 & 0.0 & 0.0 \\
filling factor & 0.045 & 1.0 & 1.0 \\
$\beta$ & 0.0 & 0.0 & 0.0 \\
He/He$_{\odot}^{d}$ & 1.0 & 1.0 & 1.1 \\
C/C$_{\odot}^{d}$ & 17.8 & 13.5 & 30 \\
N/N$_{\odot}^{d}$ & 230  & 214 & 235 \\
O/O$_{\odot}^{d}$ & 12.8 & 9.8 & 15.5 \\
$\chi^2$ & 11 & 16 & 22 \\
\end{tabular}

\medskip
{\em $^a$} Filling factor, $\alpha$ and $\beta$ allowed to vary.

{\em $^b$} Filling factor set at 1.  $\alpha$ and $\beta$ set to 0.

{\em $^c$} Uncertainties are comparable with those in Table 2.

{\em $^d$} Log of solar number abundances relative to hydrogen He:-1.0 

C: -3.45 N:-4.03 O: -3.13 (Grevesse $\&$ Noel 1993).

\end{table}

\input epsf

\setcounter{page}{12}
\begin{figure}
\vskip 1.0 in
\caption{H$\beta$ line flux as a function of time after outburst.
Squares are from Saizar et al. (1991), triangles are from
Kenyon $\&$ Wade (1986) and the diamond is from Andre\"{a}
et al. (1991).  The dotted line represents equation 1.}
\end{figure}
\newpage

\begin{figure}
\vskip 1.0 in
\caption{Top plot: Line fluxes of the observations 
and errorbars (triangles) and the best model (squares).
Bottom plot: The $\chi^2$ contribution of each line}
\end{figure}

\begin{figure}
\vskip 1.0 in
\caption{Elemental abundances of S91 (squares), A94 
(diamonds), this paper (connected pluses) and 
Grevesse \& Noel (1993) solar values (circles) for PW Vul 1984.
We have slightly offset the elemental abundances of S91 and A94 
to improve readability.} 
\end{figure}

\begin{figure}
\vskip 1.0 in
\caption{Top two plots: Electron number density and electron temperature
as a function of radius in the best fit July model.
Bottom five plots: The normalized ion concentrations for H, He, C, N and
O as a function of radius in the best fit July model.}
\end{figure}

\end{document}